\documentclass[sigconf]{acmart} 

\usepackage{graphicx}
\graphicspath{{./images/}}
\usepackage{rotating}
\usepackage{float}
\usepackage{multirow}

\AtBeginDocument{%
  \providecommand\BibTeX{{%
    \normalfont B\kern-0.5em{\scshape i\kern-0.25em b}\kern-0.8em\TeX}}}

\copyrightyear{2023}
\acmYear{2023}
\setcopyright{rightsretained}

\acmConference[CHI '23]{Proceedings of the 2023 CHI Conference on Human Factors in Computing Systems}{April 23--28, 2023}{Hamburg, Germany}
\acmBooktitle{Proceedings of the 2023 CHI Conference on Human Factors in Computing Systems (CHI '23), April 23--28, 2023, Hamburg, Germany}\acmDOI{10.1145/3544548.3581517}
\acmISBN{978-1-4503-9421-5/23/04}

\begin{document}

\title{Understanding Human Intervention in the Platform Economy: A case study of an indie food delivery service}
\author{Samantha Dalal}
\email{samantha.dalal@colorado.edu}
\orcid{0000-0002-5604-055X}
\affiliation{%
  \institution{University of Colorado Boulder}
  \streetaddress{1045 18th Street, Campus Box 315}
  \city{Boulder}
  \state{Colorado}
  \country{USA}
  \postcode{80309}
}

\author{Ngan Chiem}
\email{nchiem@princeton.edu}
\orcid{0000-0001-9831-8348}
\affiliation{%
  \institution{Princeton University}
  \streetaddress{35 Olden St}
  \city{Princeton}
  \state{New Jersey}
  \country{USA}
  \postcode{08544}
}

\author{Nikoo Karbassi}
\email{karbassi@princeton.edu}
\orcid{0000-0001-5073-6188}
\affiliation{%
  \institution{Princeton University}
  \streetaddress{35 Olden St}
  \city{Princeton}
  \state{New Jersey}
  \country{USA}
  \postcode{08544}
}

\author{Yuhan Liu}
\email{yl8744@princeton.edu}
\orcid{0000-0001-6852-6218}
\affiliation{%
  \institution{Princeton University}
  \streetaddress{35 Olden St}
  \city{Princeton}
  \state{New Jersey}
  \country{USA}
  \postcode{08544}
}

\author{Andrés Monroy-Hernández}
\email{andresmh@princeton.edu}
\orcid{0000-0003-4889-9484}
\affiliation{%
  \institution{Princeton University}
  \streetaddress{35 Olden St}
  \city{Princeton}
  \state{New Jersey}
  \country{USA}
  \postcode{08544}
}

\renewcommand{\shortauthors}{Dalal, et al.}

\begin{abstract}
This paper examines the sociotechnical infrastructure of an “indie” food delivery platform. The platform, \textit{Nosh}, provides an alternative to mainstream services, such as \textit{Doordash} and \textit{Uber Eats}, in several communities in the Western United States. We interviewed 28 stakeholders including \textit{restauranteurs}, \textit{couriers}, \textit{consumers}, and \textit{platform administrators}. Drawing on infrastructure literature, we learned that the platform is a patchwork of disparate technical systems held together by human intervention. Participants join this platform because they receive greater agency, financial security, and local support. We identify human intervention's key role in making food delivery platform users feel respected. This study provides insights into the affordances, limitations, and possibilities of food delivery platforms designed to prioritize local contexts over transnational scales.
\end{abstract}



\maketitle

\section{Introduction}
Evidence about the negative impacts of gig work platforms---digital platforms that coordinate short-term, on-demand work--- has grown in recent years. Scholars and journalists alike have documented the precarity~\cite{ravenelleSideHustleSafety2021,ma2022brush,dziezaRevoltDeliveryWorkers2021}, the invasive surveillance~\cite{ravenelle2019we}, and the psychological tolls~\cite{prasslHumansServicePromise2018,rosenblatAlgorithmicLaborInformation2016a} of mainstream gig work platforms, such as Uber, DoorDash, and Deliveroo. In response to these negative impacts, stakeholders of mainstream gig work platforms have begun to explore alternative, ``indie'' platforms. We borrow the meaning and associations of ``indie'' from music, film, games, and other industries where indie was used to identify small, independent artists supported by an enthusiastic local fanbase who operated separately from the mainstream platforms. Indie platforms are localized,
independent businesses that offer an alternative to mainstream services. These indie platforms claim to offer stakeholders more favorable terms of participation and better treatment. Despite their growing popularity, there is very little empirical HCI research exploring stakeholders' motivations for joining and their experiences within these alternative platforms.

In this paper, we are interested in indie food delivery platforms. In this context, the stakeholders we are concerned with are restaurants, couriers, and consumers. During the COVID-19 pandemic, the revenue of mainstream food delivery platforms, like DoorDash and GrubHub, more than doubled~\cite{mckinsey}. However, the increased use of these platforms was accompanied by increased dissatisfaction due to high commissions rates and fees~\cite{popper2020diners}, untenable pacing~\cite{dziezaRevoltDeliveryWorkers2021}, and one-sided communication~\cite{thompsonHiddenCostFood2020, DeliveryAppsGrubhub2021,lord2022critical,ma2022brush}. Moreover, restaurants and couriers felt that mainstream food delivery platforms did not support the human intervention needed to make a complex process like food delivery possible~\cite{figueroaEssentialUnprotectedAppbased,dziezaRevoltDeliveryWorkers2021,griswoldAreFoodDelivery2022}. Discontent with the unfavorable terms offered by mainstream platforms, coupled with the availability of white-label software\footnote{White label software is software that an organization pays for to brand as its own. This software is often sold under a subscription or transaction fee. Users usually do not know the software maker.}, facilitated the rise of indie food delivery platforms as an alternative. Despite their growing popularity, indie food delivery platforms remain understudied. 

While some restaurants, couriers, and consumers have turned to indie platforms due to their negative experiences with mainstream platforms, this model has downsides too. Unlike mainstream platforms, indie platforms are relatively small and lack a well-established customer base, making them a risky venture for restaurants and couriers. To understand why stakeholders would take on the risks associated with joining an indie platform and how they navigate the socio-technological ecosystem of an indie platform, we ask:

\begin{description}
\item [RQ1:] \textit{What \textbf{motivates} stakeholders to join an indie platform?} 
\item [RQ2:] \textit{What are the capabilities and limitations of the socio-technical \textbf{infrastructure} that powers an indie platform?} 
\end{description}

To answer our research questions, we interviewed 28 stakeholders---\textit{restauranteurs}, \textit{couriers}, \textit{customers}, and \textit{administrators}---of an indie platform called \textit{Nosh}, operating in the northern region of Colorado, in the Western United States. We asked what motivates restaurants, customers, and couriers to seek out alternative avenues for participation in the food delivery industry through joining an indie platform \textbf{and} how this indie platform functions, fails, and is repaired. We utilized thematic analysis to qualitatively code our data. 

Drawing on a conceptual lens that combines the platform economy and infrastructure studies from science and technology studies, \textbf{we find that human intervention is a commonplace and necessary part of platform-mediated food delivery}. We found that human intervention plays a crucial role in food delivery: users are willing to leave platforms that penalize interventions for alternative platforms that support them. In this context, human intervention refers to when individuals make an autonomous, ad-hoc decision that goes against algorithmically prescribed instructions, such as modifying an algorithmically recommended driving route for delivery based on road closures. Mainstream platforms tend to punish human intervention, despite its essential role in mediating a complex process such as food delivery. However, human intervention is essential to the repair work that addresses infrastructural breakdowns and is integral to the perceptions of agency and respect that stakeholders seek. This paper contributes a grounded understanding of human intervention's role in making gig work platforms viable and supporting stakeholders' perceptions of agency. Additionally, \textbf{we reflect on how business models\footnote{We use the term business model to refer to platforms' broader business practices, not only their specific plans to make a profit.} play a vital role in dictating whether socio-technical systems will support human intervention}. Drawing on our findings, we call attention to the role of business models in shaping the socio-technical systems of digital labor platforms. In particular, we highlight the need to consider the role of business models as an analytical frame when making design recommendations for the platform economy. We conclude by calling for HCI scholars seeking to improve conditions for participants in the platform economy to consider how a platform’s business model enables and constrains the impact of their design recommendations. 

\section{Literature Review}
Food delivery platforms create a two-sided market: they match the food supply from restaurants to the demand for food from consumers and coordinate food delivery through a fleet of couriers~\cite{parwezCOVID19PandemicWork2022}. During the lockdown phase of the COVID-19 pandemic, food delivery platforms provided a means for local restaurants to continue their business despite shutting down brick-and-mortar locations. Since then, food delivery platforms have cemented themselves as a permanent fixture in the restaurant industry, generating record sales and volume~\cite{pewresearch, mckinsey}. 

Despite the usefulness of food delivery platforms, they present challenges to the restauranteurs and the couriers who rely upon them. Restaurants experience financial precarity due to high commission rates, and drivers are subject to intrusive surveillance, both exacerbated during the pandemic ~\cite{thorbecke2020impact}. Researchers and journalists have identified structural flaws in food delivery and ride-hailing platforms, ranging from exploitation~\cite{van2017platform,ravenelle2019we,moore2021augmented} to algorithmic control~\cite{woodcock2020algorithmic} to discrimination~\cite{beerepoot,adermon2022gig}. While previous research has documented the harms caused by mainstream food delivery platforms, little work has been done to examine \textbf{how workers and communities have sought to address these harms by supporting local, independent platforms}. This literature review explores food delivery platforms' role in supporting the restaurant industry, problems endemic to food delivery platforms, and emergent solutions to those problems. We utilize infrastructure as a lens to add context to our findings of how these platforms function, break down, and are repaired.  
\subsection{What is Infrastructure?}
Infrastructure refers to the foundation underpinning any large-scale system; society depends on these systems to conduct everyday activities~\cite{rice2006handbook}. While infrastructure is typically conceived as physical (\textit{e.g.}, roads), information communication technologies are also infrastructures~\cite{hanseth2010design}. These information infrastructures comprise digital platforms that allow users and communities to engage with the information systems~\cite{hanseth2010design}.
Information infrastructures have become integral to how people communicate with each other and obtain goods and services in the modern economy. Infrastructures are socio-technical systems that shape and are shaped by social practices ~\cite{edwards2003infrastructure,rice2006handbook}. In other words, infrastructure \emph{``never stands apart from the people who design, maintain and use it,''}~\cite{rice2006handbook}. Thus, infrastructures are value-laden -- they are imbued with the values of their designers~\cite{star1996steps,rice2006handbook}. When designers' values do not align with users' values, friction occurs which can result in breakdowns. Breakdowns occur when users encounter ``insurmountable incongruence between the expected infrastructure service and its actual or perceived behavior.''~\cite{pipek2009infrastructuring}. For example, couriers working for food delivery apps can experience breakdowns when the routing instructions given by the platform tell them to take a closed road due to poor weather conditions. Examining when infrastructural breakdowns occur allows us to surface the values of designers and users~\cite{dye2019if}.

To mitigate breakdowns, infrastructures should be designed with users and integrated into existing structures~\cite{rice2006handbook,lee2006human,pipek2009infrastructuring}. However, this requires the acknowledgment of local, end-user expertise~\cite{bowker2009toward}. In the context of the platform economy, where designers are building to achieve scale and universality of their product, the development of flexible systems that incorporate end-user expertise is complicated~\cite{qadri2022seeing}. Qadri and D'Iganazio posit that platforms have a fundamentally different ``vision''~\cite{qadri2022seeing} of the landscape of work from drivers, making it difficult for them to see and account for end-user expertise~\cite{qadri2022seeing}. The infrastructure of the platform economy ``mediates between the local and global''~\cite{lee2006human}: UberEats is an infrastructure that translates between the global concern for on-demand food delivery and the local concern of how traffic-snarled streets can be navigated as efficiently as possible to complete a delivery order. However, as Qadri demonstrates in her ethnography of delivery drivers in Indonesia~\cite{qadriPlatformWorkersInfrastructures2021}, the translations from local to global in the context of the platform economy are slipshod at best. End-users often develop their own human infrastructures---\emph{``arrangements of organizations and actors that must be brought into alignment for work to be accomplished''}~\cite{lee2006human}---to make up for the gaps in translation~\cite{chandra2017market,gray2019ghost,lee2006human}. Human infrastructures play a crucial role in supporting information infrastructures~\cite{lee2006human}. For example, communities come together to maintain and care for networked infrastructures~\cite{dye2019if}, geographically isolated gig workers develop communication networks to make micro-task work tenable~\cite{gray2019ghost}, and delivery drivers build coalitions to maintain rest stops for other drivers~\cite{qadri2022seeing}. Infrastructure literature has thoroughly documented the importance of human infrastructure in supporting large-scale systems~\cite{lee2006human,pipek2009infrastructuring,rice2006handbook}. Despite their importance, human infrastructures are often invisible by nature~\cite{lee2006human}.

In this paper, we use infrastructure as a lens to look beyond the surface-level system functioning to see the power relations embedded in the values that motivate people's labor on these platforms. We specifically focus on the interaction between human infrastructures and the infrastructures of the platforms they support. In doing so, we can better understand the entanglement between local practices and global technologies and why, despite their importance, these local practices remain invisible to some platforms. We adopt Jackson and colleagues' ``broken-world''~\cite{jackson2012repair} thinking to center the ongoing labor involved in maintaining gig work platforms, the values imbued within the platform economy, and the underlying power structures. In the following sections, we examine the infrastructure of platform-mediated food delivery to understand the values imbued within mainstream food delivery platforms, how those values create friction that leads to infrastructural breakdowns, and how end-users engage in repair work to make platforms viable.

\subsection{Food Delivery Platforms and Their Problems}
During the COVID-19 pandemic, demand for the largest food delivery services like Grubhub and UberEats grew as delivery became integral to restaurants' business models~\cite{smithPeopleHaveEat}. However, researchers have identified several harms specific to food delivery platforms, including exacerbating workers’ and restaurants’ economic insecurity and pervasive algorithmic control~\cite{parwezCOVID19PandemicWork2022,frankeConnectingEdgeCycles2021,huangRidersStormAmplified2022,woodcock2020algorithmic}. Harms often occur as a by-product of infrastructural breakdowns and thus can be used as an entry point to surface underlying values of platform designers~\cite{Semaan2019,friedmanValueSensitiveDesign2013a}. In the following sections, we identify how value-misalignment between designers and users manifests in food delivery platforms and the negative impacts this has on both couriers and restaurants. 
\subsubsection{Value Misalignment Between Designers and Users of Food Delivery Platforms}
Food delivery platforms are designed by software engineers and funded by venture capitalists and are thus imbued with their values~\cite{zuboff2015big}. Food delivery platforms, like DoorDash, are designed by software engineering teams who prioritize high-speed, low-cost delivery by assigning orders to gig workers available nearby regardless of gig workers' preferences~\cite{weinsteinUsingMLOptimization2021}. Couriers are constantly pushed to deliver more, faster: the expected speed of their next delivery is determined by the speed they were able to achieve with past deliveries ~\cite{dziezaRevoltDeliveryWorkers2021}. This has sometimes led to couriers getting into fatal traffic accidents as they speedily navigate roads on electric bicycles~\cite{figueroaEssentialUnprotectedAppbased}. Platform interfaces are artifacts that reveal the underlying values of their designers; there are few, if any, features on the couriers' mobile application interface that allow them to articulate their preferences to the platform~\cite{liBottomUp}. Numerous infrastructure scholars have pointed to the importance of integrating user input into the design process to ``manage issues of complexity and standardized interfaces between new and existing work infrastructure''~\cite{pipek2009infrastructuring}, yet in most food delivery platforms there are few avenues for this communication to take place. When designers' values fail to align with and account for users' values, such as couriers, problems arise, resulting in users being harmed~\cite{Semaan2019,friedmanValueSensitiveDesign2013a}. 
\subsubsection{Problems that couriers experience in the food delivery economy}
Unlike traditional food delivery services (\textit{e.g.}, Pizza Hut phone order delivery), platform delivery services heavily incorporate algorithms into their operations to not only convey logistical information to couriers about their tasks, but also to track their performance~\cite{goods2019your}. This omnipresent surveillance allows companies like UberEats to track the rate at which food couriers accept or deny delivery requests and their customer ratings to compile a performance rating of each worker~\cite{veenPlatformCapitalAppetiteControl2020}. Mainstream food delivery platform designers must rely on this constant surveillance to enforce the quality of service provided~\cite{rahmanInvisibleCageWorkers2021a}. However, constant surveillance and uncertainty about how they are managed causes emotional and psychological stress for couriers~\cite{caloTAKINGECONOMYUBER2017,woodcock2020algorithmic,rosenblatAlgorithmicLaborInformation2016a}. Unpredictability of working conditions and opacity around algorithmically-determined payments creates additional precarity for workers~\cite{goods2019your,calacciBargainingBlackBox2022}. 

Couriers participating in the platform food delivery industry take on financial risk as part of their work. For example, shifting consumer demands, poor weather conditions, and irregular unpaid wait times for restaurant food preparations are common occurrences that can bring couriers’ earnings to below minimum wage~\cite{goods2019your}. Opaque algorithmic management also causes wage insecurity. Food delivery platforms categorize couriers as independent contractors, meaning they must demonstrate that workers have meaningful control over their decisions in the workplace~\cite{dubalAlgorithmicWageDiscrimination2023}. To do so, platforms rely on complex algorithmic systems to create incentive systems that act as a proxy for control but maintain a level of inscrutability that allows them to eschew the responsibilities associated with being an employer~\cite{rosenblatAlgorithmicLaborInformation2016a,krzywdzinskiAutomationGamificationForms2021}. This results in wage calculations that are unpredictable and highly variable furthering causing financial precarity for workers~\cite{dubalAlgorithmicWageDiscrimination2023}. 

While platform designers are not intentionally trying to develop technologies that harm couriers, the intentional \textit{design choices} they make result in harm regardless. Platform designers value scale and thus embrace abstraction of local contexts: they embrace ``scale-thinking''~\cite{hanna2020against}. This can lead to a failure to consider how their algorithms, such as those that continually increase the expected speed of deliveries, function on the ground where couriers encounter physical barriers and must slow down to be safe~\cite{dziezaRevoltDeliveryWorkers2021}. Business values play a crucial role in dictating what types of interactions are supported by the infrastructure~\cite{shestakofsky2020making}. Value-misalignment between designers and users of food delivery platforms also extends to restaurants.

\subsubsection{Problems that restaurants experience in the food delivery economy}
After indoor dining was restricted during the pandemic, many restaurants signed up for food delivery services as an alternative revenue stream. However, the mainstream platforms’ high commission rates cut into the earnings of already struggling local businesses~\cite{saxenaDeliveryAppsAre2019}. Restaurants hesitated to raise prices because doing so may drive away price-sensitive consumers~\cite{DeliveryAppsGrubhub2021, thompsonHiddenCostFood2020}. This put participating restaurants in a situation where they had to take a hit on their already thin profit margins to avoid driving away consumers. These negative impacts have driven some restaurants and couriers to explore alternative options, such as indie platforms, for participating in the food delivery platform economy. 


\begin{table*}[h]
\begin{tabular}{lp{7cm}p{6.5cm}}
\toprule
\textbf{Service Provided } & \textbf{Description} & \textbf{Example Platforms} \\ \midrule
Website Builders & Provide web presence for restaurants & Bentobox, SquareSpace, Wordpress, Wix   \\ \midrule
Point of Sale (POS) & Process online transactions  &  Square, Toast, Aloha, Clover, Lightspeed \\ \midrule
White Label Ordering  & Enable direct ordering on website   & ChowNow, Bbot, 9Fold, GoParrot, Lunch Box  \\ \midrule
Order Integrators & Integrates incoming orders from multiple services & Ordermark, Chowly, Cuboh, Itsacheckmate, Otter \\ \midrule
Delivery Services & Provide delivery services from restaurant to consumer & Postmates, Relay, JOLT, Habitat, DoorDash Drive \\ \bottomrule
\end{tabular}
\caption{Software and Services in the Unbundling Economy}
\label{tab:unbundling}
\end{table*}

\subsection{The Emergence of ``Indie'' Platforms as a Response to Infrastructural Breakdowns}
Small, localized, and community-centric alternative gig platforms have surfaced to fill the gaps left by mainstream services~\cite{atkinsonMoreJobFood2021}. These ``indie'' platforms claim to offer restaurants, drivers, and customers a service that charges lower commissions, cares about the stakeholders involved, and is on the side of the community ~\cite{atkinsonMoreJobFood2021, schneiderExitCommunityStrategies2020,WhyAreRestaurants2021,DeliveryCoopsProvide}. However, there are also drawbacks to indie platforms. They usually do not have as large of a customer base or as sophisticated of technology as mainstream platforms~\cite{DeliveryAppsGrubhub2021}. Research into these new models has been limited, often only focusing on one subset that function as co-operatives~\cite{bundersFeasibilityPlatformCooperatives2022a,scholz2016and,schor2021after}. Our research seeks to expand the conversation around these alternative designs by applying the CSCW and HCI infrastructure literature to understand why these platforms emerge, how they function, and their limitations and advantages. 
\subsubsection{Indie platforms rely on the ``unbundling'' economy for technical infrastructure}
Indie platforms draw upon the ``unbundling'' economy to build their own services from the bottom up. This economy provides software and services for individual elements of the food delivery process: point of sales systems, order aggregation services, and last mile delivery fleets (see Table~\ref{tab:unbundling} below)~\cite{wangCommissionfreeOnlineOrders2020,wangArmingRebelsFood2021}. These piecemeal systems that attempt to integrate disparate software and services often break down due to incompatibility and difficulty of use. However, prior literature in HCI provides guidance on how users engage in \textit{bricolage} to make piecemeal systems work when mainstream offerings breakdown~\cite{vallgaarda2015interaction,jackson2014RethinkingRepair}. Through bricolage, users extend and repair existing systems through human labor to build an infrastructure that better meets their needs~\cite{vallgaarda2015interaction,jackson2014RethinkingRepair}. Firms recombine existing resources (such as those available through the ``unbundling economy'') to create something out of nothing~\cite{baker2005creating}. In doing so, indie food delivery platforms refuse to accept the limitations of the dominant definitions of resources required to compete in the platform economy. Whereas typical firms in the platform economy are well-funded technology firms staffed with engineers, indie food delivery platforms are small, grassroots-funded operations with some technical expertise. Indie platforms can operate in the same markets as mainstream platforms because they engage in the creative practice of bricolage~\cite{baker2005creating}. However, the practice of bricolage necessitates human labor and creativity~\cite{baker2005creating,jackson2014RethinkingRepair,vallgaarda2015interaction}.

\subsubsection{Human labor plays a key role in supporting indie platforms}
Indie platforms build their services by integrating disparate pieces of software into one technology stack, usually without having a fully-staffed software engineer team on hand~\cite{DeliveryCoopsProvide}. In this low-resource environment, technological breakdowns are inevitable~\cite{DeliveryCoopsProvide}. Despite this, indie platforms have continued to survive and grow in prominence because of their small-scale, localized nature and human-centric design~\cite{atkinsonMoreJobFood2021,DeliveryCoopsProvide,holtzCanCoOpsFix2022}. However, little empirical work has been done to understand how indie platforms overcome technological shortcomings. We draw upon the concept of bricolage to better understand how users of indie platforms overcome failures in their technology stack to create a cohesive food delivery service within a low-resource environment~\cite{baker2005creating}.   

This paper draws upon infrastructure literature and applies it to the context of the food delivery platform economy to better understand why users join indie platforms and how these platforms continue to survive despite their lower levels of technological sophistication. We identify how stakeholders surface the values embedded in the infrastructures of mainstream platforms, which become evident during breakdowns, and how value-misalignment motivates them to join indie platforms. We examine how top-down construction of infrastructures fails to account for local contexts and highlight the importance of designing to incentivize human intervention in the face of breakdowns. This work expands on design literature in the platform economy by demonstrating how HCI scholars can develop systems that align with local communities' needs and mesh with the economic incentives of the business model. 

\section{Background on mainstream and indie food delivery platforms}
In this paper, we differentiate between mainstream and indie food delivery platforms. Here we describe the characteristics of each to highlight the key differences between the two models. We then provide additional details about the empirical setting where we conducted the research.

\subsection{Mainstream food delivery platforms}
Mainstream platforms have business models predicated on \textit{market capture}, or the monopolization of a given market. Mainstream platforms have access to large pools of venture capital, enabling them to operate at a loss ~\cite{DoorDashSharesSink, GrubHubOrdersKeep}. Thus, mainstream platforms are not solely reliant on commissions from orders for profit; rather, they simply need to demonstrate growth or the potential for growth to continue to receive funding. This means that mainstream platforms' success is not strictly predicated on the success of the restaurants and drivers participating in them. This business strategy informs both the technical design of their systems and therefore shapes the nature of interactions that stakeholders have with each other and the platform. 

\begin{figure*}[h]
\caption{Difference in financial flows between mainstream platforms and Nosh}
\label{fig:financialflows}
\includegraphics[width=\textwidth]{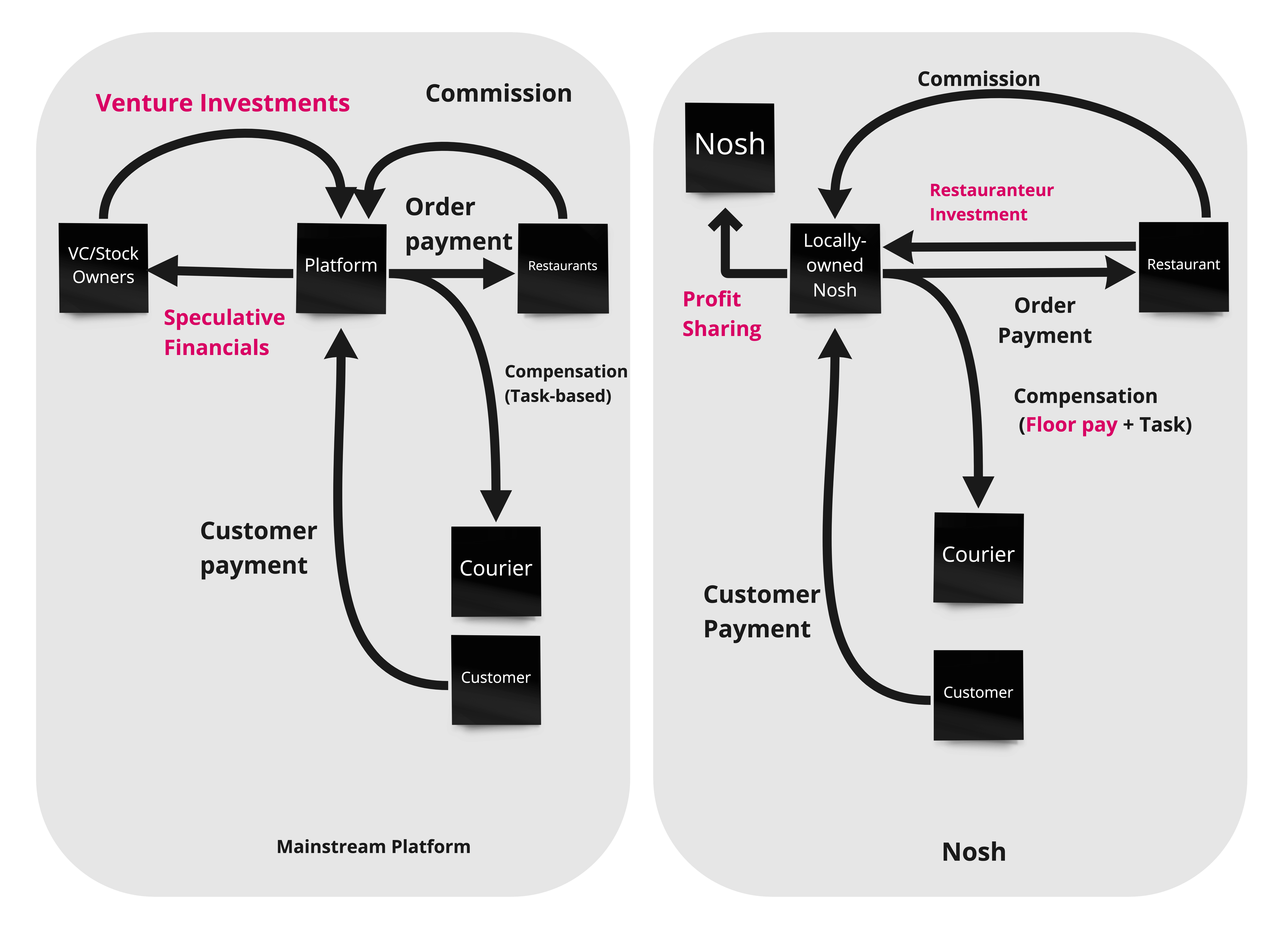}
\end{figure*}

\subsection{Nosh: an indie food delivery platform}
\subsubsection{Organizational and financial structure} 
In contrast to mainstream platforms, Nosh's business model is rooted in the local communities in which it operates. 
Nosh operates as a ``hub and spoke'' business model. This means that the parent company, Nosh, provides the technology, brand, and initial seed capital for the smaller, local Nosh ``spokes.'' When Nosh comes to a new community, they establish a local Nosh delivery service (\textit{e.g.}, similar to how franchises of McDonald's operate). Restaurants that join Nosh tend to be locally owned and operated. At the time of our study, no national chains, like McDonald's or Chipotle, were on the Nosh platform. Each local Nosh delivery service is funded by a coalition of local restaurants that invest seed money into their spoke. These ``investor'' restaurants get voting power over operational decisions in their locale. Investor restaurants pay a buy-in fee of approximately \$5,000 which entitles them to a say over operational decisions and a fixed commission rate of 15\%. Restaurants wanting to join Nosh but not wanting to be investors can join the platform without paying a buy-in fee. However, these \textit {member} restaurants do not get a direct say over operational decisions for their locale and pay a slightly higher commission rate, typically around 20\%.

The parent company helps each local spoke get started by hiring dispatchers and drivers from the community. The parent company also provides the instruction and technical infrastructure needed to run the application. In return, each local Nosh spoke shares a percentage of its profit with the parent company. Once dispatchers and drivers have been hired, the investor restaurants begin delivering through the Nosh application. Because Nosh does not have access to venture capital, the platform relies on commissions from their orders and investment from local restaurants. In other words, Nosh's success as a business model is predicated on its stakeholders' success. In Figure~\ref{fig:financialflows} we illustrate how the economic structure of Nosh differs from that of mainstream platforms, particularly how monetary incentives shape relationships between stakeholders and the platform.

Mainstream models rely upon investments from venture capitalists and individual shareholders to finance their operations. This makes mainstream platforms less dependent upon patronage from customers and restaurants as they can access alternative cash flows. Nosh's model differs from mainstream platforms in terms of economic incentives because the primary investors in the platform are the restaurants themselves. As investors/owners of each local Nosh hub, restaurants contribute to operational decisions and are the platform's primary source of cash flow. The platform's viability is thus tied to stakeholders’ fiscal well-being. Through this model, Nosh is \textbf{beholden} to its stakeholders in ways that mainstream platforms are not. This impacts how the Nosh platform operates, especially in times of breakdown.

\subsubsection{Nosh relies on a patchwork of technical systems to function}
Food delivery relies upon multiple parties and disparate pieces of software being able to communicate with each other. Nosh uses a suite of proprietary and software-as-a-service technologies to run their business that include:
\begin{itemize}
\item \textit{DataDreamers} for delivery logistics. Although Nosh reported moving to another company's software.
\item \textit{WhenIWork} for courier scheduling and chatting with couriers.  
\item \textit{Zendesk} for customer support.
\item \textit{Slack} for internal communications between dispatchers. 
\item \textit{Google Suite} for generating earning reports for restaurants.
\end{itemize}

While these pieces of software are more affordable than a vertical integration model, breakdowns and confusion often occur, which can frustrate stakeholders.

\subsection{Empirical Setting} 
We conducted an interview study with four stakeholder groups participating in Nosh: restauranteurs, couriers, platform administrators, and customers, to understand what motivates people to join an indie platform. During our interviews, Nosh operated in Fort Collins and Boulder, Colorado. The Fort Collins region encompassed multiple smaller cities whereas the Boulder region contained only one major metropolitan area. We conducted interviews from August to December of 2021. The region is also served by mainstream platforms such as GrubHub, Uber Eats, DoorDash, and Postmates. 

\section{Methods}
Our study aims to better understand the workings of an indie food delivery operation. We conducted 28 interviews with four stakeholder groups participating in Nosh: restaurants, couriers, customers, and platform administrators. Before conducting the interviews, our protocol and procedures were certified by our institutional review boards. The interview protocols differed for each stakeholder group, but all sought to surface why the subject chose to use Nosh and how they participated in Nosh. We include the interview protocols for each stakeholder group in the appendix. Below we provide more detail on the participants, interview protocol, and data analysis.

\subsection{Participant Recruitment}
We contacted the founder of Nosh who helped us distribute a call for participation to an email list sent out to all restaurants, drivers, platform administrators, and customers who used Nosh.

In total, we conducted 28 interviews across four stakeholder groups: restauranteurs (n=7), couriers (n=6), customers (n=10), and platform administrators (n=5). Most restauranteurs worked for or owned fast-casual businesses (R1, R2, R4, R5, R6, R7). We spoke to platform administrators who did back-end work (P2), customer-facing work (P1, P5), and managerial work (P3, P4). The drivers we spoke to were evenly split among part-time (D1, D2, D4) and full-time drivers (D3, D5, D6). We  break down our participant demographics in Table~\ref{table:stakeholderlist}.

\begin{table}[h]
\begin{tabular}{ll}
\toprule
\textbf{Participant} & \textbf{Occupation Description}    \\
\midrule
\multicolumn{2}{c}{\textit{Restauranteurs}} \\
\hline
R1                   & COO - Chicken Wing Restaurant\\

R2                   & Owner - Tea Shop                \\

R3                   & Owner - Seafood Fine Dining     \\

R4                   & Owner - Wood Fired Pizza        \\

R5                   & Owner - Fast-Casual Take Out    \\

R6                   & Owner - Chicken Wing Restaurant \\

R7                   & Owner - Burger Restaurant       \\
\hline
\multicolumn{2}{c}{\textit{Platform Administrators}}                       \\
\hline

P1                   & Dispatcher                                    \\

P2                   & Data Analyst                                  \\

P3                   & Co-Director of Operations                     \\

P4                   & Co-Director of Operations                     \\

P5                   & Dispatcher                                    \\
\hline
\multicolumn{2}{c}{\textit{Couriers}}                       \\
\hline

D1                   & Driver - Part Time                                    \\

D2                   & Driver - Part Time                                  \\

D3                   & Driver - Full Time                     \\

D4                   & Driver - Part Time                     \\

D5                   & Driver - Full Time                                    \\

D6                   & Driver - Full Time                                    \\
\hline
\multicolumn{2}{c}{\textit{Customers}}                       \\
\hline

C1                   & Health Care Professional                                    \\

C2                   & Office Worker                                  \\

C3                   & Office Worker                     \\

C4                   & Homemaker                     \\

C5                   & Office Worker                                    \\

C6                   & Office Worker                                    \\

C7                   & Office Worker                                    \\

C8                   & Undisclosed                                  \\

C9                   & Office Worker                     \\

C10                   & Undisclosed                    \\
\bottomrule
\end{tabular}
\caption{List of participants broken down by the four stakeholder categories: restauranteurs, platform administrators, couriers, and customers}
\label{table:stakeholderlist}
\end{table}

\subsection{Interview Protocol}
The first, second, and third authors conducted one-hour interviews via Zoom. We informed participants of their right to skip  questions or stop the process at any time. We obtained their verbal consent for recording the interview and allowed them to sign up for a follow-up survey for future research. We transcribed the interviews and anonymized the transcripts. We compensated participants for their time with a \$20 Amazon gift card.

Our interviews followed semi-structured scripts based on the participant's group affiliation\footnote{see the Appendix for interview protocols}. These scripts focused on surfacing key motivations for participating in Nosh and understanding how participants interacted with the platform. For example, depending on their role, we discussed the pros and cons of delivering for Nosh (drivers), collaborating with Nosh (restaurateurs), ordering from Nosh (customers), and working for Nosh (dispatchers). Specifically, for restaurants, couriers, and customers we first asked them about what food delivery platforms they use regularly and why they chose to use Nosh. We then explored how these stakeholders interacted with both Nosh's digital interface and other stakeholders involved in the food delivery process, paying special attention to how they handled breakdowns in the food delivery process both within the context of mainstream platforms and Nosh. This set of questions aimed to better understand how these stakeholders handled and repaired infrastructural breakdowns. We concluded interviews with these stakeholders by asking them to describe what a co-op is---because Nosh describes itself as a co-op, even though it is not legally incorporated as one---and what the differences were between this business model and mainstream platform business models. For the platform administrators, we started by asking them to describe their roles and the suite of tools they used to do their jobs. We interviewed the platform administrators by asking them about their interactions with other stakeholders and the community where they operated to better understand the social interactions necessary to make platform-mediated food delivery functional. The first, second, and third authors manually corrected the automatically-generated transcriptions provided by Zoom by cross-referencing them with the audio recordings. 
\subsection{Data Analysis}
After transcribing each interview, the first, second, and third authors coded the contents using thematic analysis. To standardize the analysis between three people, we conducted preliminary coding where the researchers independently analyzed the same batch of interviews. We then convened afterward to compare and discuss our findings. We also conducted card-sorting exercises to arrive at the final codes. Later, we compiled repeating themes in a centralized codebook, which we then used to code the remainder of the interviews.

We frame our findings around two key themes: \textbf{(1) Users' motivations for participating in indie platforms and (2) How users and technology come together to make indie platforms work}. In Section~\ref{sec:motivations} we first identify the essential roles of agency and financial security in motivating restaurants, couriers, and customers to join Nosh. We highlight how the presence of human intervention in the face of breakdowns is a differentiating factor between mainstream platforms and Nosh. In Section~\ref{sec:patchwork} we then examine how users interact with Nosh's technical systems, paying special attention to the key role of platform administrators in serving as buffers between users and the technical systems.


\section{Why do stakeholders join an indie platform? More agency and financial security}\label{sec:motivations}

Restaurateurs and couriers joined the indie platform because of a desire for more agency and financial security. Similarly, customers joined because they perceived the indie platform as more ethical, precisely for the same reasons: more agency and financial security for restaurateurs and couriers. See Table~\ref{tab:motivations} for a synthesis of these motivations.
\subsection{More agency motivates stakeholders to join}
When participating in mainstream platforms, stakeholders (couriers, restauranteurs, and customers) noted that they must fit themselves into the logics of the platforms with little room for negotiation. Processes for participation, operation, and dispute resolution were predetermined by mainstream platforms. This limited stakeholders' ability to exert meaningful control over their participation, advocate for themselves, and come to resolutions they perceived to be fair. In contrast, stakeholders felt that their agency was preserved when using Nosh, which motivated their adoption of the service. Stakeholders’ perception of agency was closely aligned with the presence of human intervention: the ability to interact, reason, and negotiate with a human representative from Nosh gave them a sense of agency that they lacked when participating in mainstream platforms. We demonstrate how agency was defined and identified by restaurants, couriers, and customers below.

\begin{table*}[h]
\centering
\begin{minipage}{1\textwidth}
  \centering
    \begin{tabular}{ll}
        \toprule \textbf{Stakeholder Group} & \textbf{Motivations for Participating In Nosh}  \\ \midrule
        Restaurants & Lower commissions, Increased operational oversight, Dislike of mainstream platforms  \\
        Couriers & Floor pay, Guaranteed shifts, Increased autonomy \\
        Customers  & Support local, Dislike of mainstream platforms, Better customer service \\ \bottomrule
    \end{tabular}
    \caption{Restaurant, Courier, and Customer Motivations for Participating in Nosh}
    \label{tab:motivations}
\end{minipage}
\end{table*}

\subsubsection{Restauranteurs' agency through transparency and operational input}
Negative experiences with mainstream platforms, such as encounters with dark patterns, drove restauranteurs to seek alternatives. For example, restauranteurs pointed to non-consensual listings of their business on mainstream platforms as a violation of their agency: 
\begin{quote}\textit{
     ``Grubhub will find our menu and put it online without telling us ... and then so you'll get a phone call from somebody, and then they'll place an online order to pick up. And when they pick up, it's a Grubhub driver. What they do later is contact you and say, hey look you had X amount of deliveries [through] Grubhub, and you know, do you want to sign up for Grubhub?'' (R5)
}\end{quote}

In contrast, because each local hub of Nosh operates as a cooperative run by local restauranteurs, restauranteurs had a direct role in influencing business operations: 
\begin{quote}\textit{
    ``there's a lot of people, giving input and their thoughts, [Nosh] is truly a cooperative where you know matters of importance are voted on and discussed openly. It's very transparent operation from Nosh, the other services, you know no transparency [is] there you know it's just a plug and play ...  I don't have any say on what happens once I sign up.'' (R6)
}\end{quote}
Having a say over and transparency into operations meant that, in Nosh, restaurants were partners with agency to negotiate their participation. Restauranteurs felt that their ability to negotiate terms of participation with Nosh led to a more equitable relationship than the one they had with mainstream platforms.

\subsubsection{Courier's agency through respect for discretionary decisions}
In mainstream platforms, couriers get their work allocated through an algorithmic system. Also, mainstream platforms tend to punish couriers for deviating from the prescribed instructions even if they had good reason to do so. \begin{quote}\textit{
     ``One thing I also like with Nosh versus DoorDash – Nosh does not track my movements and my location in a way that ... If I'm not heading to the restaurant and if I'm biding my own time, it's not going to tell me to turn around or else I'll lose the order. That's something that I remember with DoorDash, that even if there was a large wait time and I didn't head in the direction of the restaurant first, the app would track that and say, 'Are you heading to the restaurant? Looks like you're not heading in the direction of the  restaurant, if you're not going to be heading there in the next couple minutes, we're going to reassign your order to somebody else'.'' (D2)
}\end{quote}
When couriers worked for Nosh, they exercised their best judgment, even if it went against what the dispatching algorithm prescribed, and not be punished for it. The ability to exert control over their work without fear of retribution meant that couriers had more agency when working with Nosh.  Several of them pointed out that they felt comfortable deviating from the instructions of the dispatching algorithm because they knew there was always a human dispatcher supervising the system. 
\begin{quote}\textit{
    ``overall, our system is much more fair than other services that I worked for because you have that human aspect of problem-solving.'' (D5)
}\end{quote}

Platform administrators could be spoken to and rationalized with, unlike an algorithm. 

\subsubsection{Customers agency through an ability to get a human for support}
Customers discussed their dispute resolution experience with mainstream platforms, emphasizing the one-sided nature of the process with mainstream platforms. \begin{quote}\textit{''One of the times I ordered formula, [Postmates] ... delivered the wrong kind. So I reached out, and they told me I could return the item. I mean, I get in my car, go return the item, and get what I wanted, and then they gave me a credit. Which I was, like, all right fine whatever, like, that's very irritating that, like, you should really just send somebody out with the correct thing. And then they refunded my account, but it was a credit on Postmates and then when I went to use it had been ... a while my credit expired ... I ended up ... not getting that money at all.'' (C2)
 }\end{quote}Here C2 details the frustration that many customers felt with mainstream platforms: they were not given the option to negotiate what kind of resolution they wanted. In contrast, with Nosh, customers could contact a real, local person, for help. Customers feel like they had agency over the situation because they could speak with a human and have a say in resolving problems. Customers pointed to the fact that platform administrators always asked them ``what can we do to make this right'' as evidence of their ability to negotiate the outcome. 

 In sum, stakeholders' sense of agency was intimately linked to the presence of human intervention during the course of food delivery. Platform administrators acted as intermediaries between stakeholders and the technical infrastructure of Nosh. Stakeholders could interact and negotiate with platform administrators in a way they could not with the automated interfaces of mainstream platforms. The ability to engage in two-sided interactions was critical to stakeholders’ agency.
\subsection{Greater financial security motivates stakeholders to join}
Mainstream platforms dictate the financial costs and benefits of participation. Stakeholders reported that this one-sided negotiation infringed on their ability to achieve financial security in the already low-margin business. With Nosh, restauranteurs reported having lower commission rates and peace of mind knowing that the platform's financial success was tied to theirs. Couriers reported that they had transparency into the fees Nosh extracted from each order, plus the satisfaction of guaranteed floor pay\footnote{Floor pay is a minimum payment per shift regardless of the number of orders based on the local minimum wage. If couriers did not meet local hourly minimum wages while working, and were not ``gaming'' the system by simply rejecting orders assigned to them, their wage was supplemented by Nosh so that it was equal to the local hourly minimum wage.}, reducing the precarity endemic to independent contractor jobs. Below, we demonstrate how restaurants and couriers attained higher degrees of financial security with Nosh and how customers perceived this as an indicator that Nosh was a more ethical option for food delivery. 

\subsubsection{Restauranteurs' financial security through lower fees and co-ownership}
Restauranteurs reported being too small to negotiate the on average 30\% commission that mainstream platforms charged them. Thus, restauranteurs with already thin profit margins had to raise menu prices to make up for the high commissions charged by mainstream platforms, which they worried would affect their customer base: 
\begin{quote}\textit{
    ``Commission rates are important  ...  and we do raise our prices significantly on delivery platforms to recover a portion of that commission'' (R7). 
}\end{quote}
According to conversations the research team had with Nosh administrators, Nosh offered a more financially sound alternative because they charged a 15-20\% commission rate.

Additionally, since each Nosh hub is owned and operated by local restauranteurs, the financial success of the platform is tied to that of the restaurant owners. Thus, restauranteurs had the added security of knowing that,  \textit{``... it's not just Nosh's financial success. They're also looking out for each individual  restauranteur's success because \textit{\textbf{[Nosh is]}} those individual restauranteurs``} (R2, emphasis added).

\subsubsection{Couriers' financial security through floor pay and guaranteed shifts}
Couriers reported that, in mainstream platforms, the work was not as flexible as platforms claim because they cannot \textit{get} work when they log onto the platform. 
\begin{quote}\textit{
     ``One of the things with the scheduling and the shifts that they [Nosh] offer is there as a preventative measure to not over-saturate the market with drivers because when you sign on for, like, DoorDash or, like, Grubhub, or any of those, there could be hundreds and hundreds of drivers ... with Nosh, you know, you can kind of be the big fish in a small pond. If you work hard, and you're efficient and fast, you can make all the money. So there's less saturation, I guess, is probably the main thing I like about it.'' (D3)
}\end{quote}
This courier explicitly calls out Nosh's scheduling system as a key element in promoting financial security. Couriers we interviewed characterized Nosh's hybrid scheduling-open shift system as being \textit{both} more flexible and secure than the ``log on and drive'' model of mainstream platforms. Nosh's hybrid scheduling-open shift system allows couriers to submit their preferred schedules a week in advance and then assigns couriers shifts based on their availability and preferences. For example, a courier who wanted to work on Tuesday evenings might be assigned the Tuesday dinner rush shift. Couriers could drop shifts, without penalty, after they were assigned by notifying Nosh dispatchers and these shifts would become available in the open shifts tab of the courier's mobile application. This scheduling model and Nosh's guaranteed floor pay provided couriers with a sense of financial security that motivated their participation in Nosh.

Increased agency and financial security made stakeholders feel respected by Nosh: the platform centered on food delivery and the stakeholders upholding it. As one restaurant put it: \textit{“DoorDash and Grubhub are fantastic technology companies that deliver food. Nosh is a fantastic food delivery company that has some technology.”} (R6). This restauranteur  highlights the primary trade-off that stakeholders who participate in Nosh make: access to top-of-the-line technology to facilitate food delivery versus access to more human-centered services that promote agency and respect. Nosh's respect for agency and financial security of restaurants and couriers motivated customers to join as they sought more ethical alternatives for food delivery platforms.

\subsubsection{Perception of ethicality motivates customers}
 Customers felt that how Nosh handled food delivery's financial aspect was more ethical than mainstream platforms.
 Nosh gave couriers and restauranteurs a degree of financial security that they could not achieve when working for mainstream platforms. 

Customers articulated feeling strongly about how both restauranteurs and couriers were treated financially by food delivery companies: 
\begin{quote}\textit{
    ``I mean, I read the news article, you know, in the Times about how Grubhub and DoorDash are just ripping off restaurants, like, it's not okay to take advantage of people during a pandemic, so I prefer not to use companies that do that.'' (C3)
}\end{quote}

Most customers we interviewed described being avid supporters of local restaurants before the COVID-19 pandemic. When the pandemic began, customers sought out ways to patronize local restaurants in an effort to support their community: \textit{``I love that [Nosh] is local, so that's my biggest thing ... it has a lot of, like, my favorite local restauranteurs''} (C4). 

Customers noted that even though other food delivery platforms, like DoorDash and UberEats, also had some local restaurants listed, they preferred ordering from Nosh because it was \textit{``just a little bit more ethical''} (C1). For customers, ethical meant that the platform treated both couriers and restauranteurs respectfully by not engaging in exploitative financial practices.



\section{What Makes Nosh Work?: A patchwork of technical systems held together by human intervention}\label{sec:patchwork}

Nosh is a patchwork of technical systems held together by human intervention. Nosh relies upon multiple disparate pieces of software to operate. Breakdowns in their patchwork technology stack are commonplace and must be repaired through human intervention. This human intervention is what compels stakeholders to participate in Nosh. However, human intervention requires deep local knowledge, takes time, and costs money. In this section, we outline the contours of Nosh’s technology stack and identify how stakeholders engage in repair work to address gaps and breakdowns, paying special attention to the role of platform administrators. We summarize what technologies each stakeholder interacts with and how they engage in repair work in Table~\ref{tab:repairwork}.

\subsection{Stakeholders struggle with Nosh's patchwork of technical systems}
Restauranteurs we interviewed complained that Nosh's ordering system was clunky and required them to operate a separate order aggregation system and a tablet dedicated solely to Nosh's orders. Mainstream food delivery platforms can integrate directly with most restaurants' point-of-sale software and order aggregation systems, reducing friction in the system. Couriers complained that as a result of the complicated order aggregation system, restaurants often forgot important steps such as starting a countdown timer that couriers relied on to estimate when the order would be ready for pickup: 
\begin{quote}\textit{
    ``it's hard you know because everybody wants to point fingers, so it's hard ... to say, like, if they're actually just, like, \textbf{not able to understand how the app works,} or they just didn't start the order because they, you know, were lazy or busy or whatever. And they're just like oh like, it wasn't actually my fault, like \textbf{the App is bad}.'' (D3, emphasis added)
}\end{quote}

Nosh’s complex technical interface and the inability to integrate it into existing infrastructure resulted in frequent breakdowns that impacted restauranteurs and couriers. Despite their familiarity with the system, platform administrators also articulated difficulty navigating the patchwork technical stack Nosh used. 

During the course of a shift, platform administrators said they used, on average, four different platforms simultaneously. One platform administrator describes their shift as follows: \begin{quote}\textit{
    ``... it's a nightmare, we need to ... streamline it because my computer just wants to it's like about to  start on fire, because I've got so many things running yeah so so slow, because I always have a million things open''(P3)
}\end{quote} 
Jumping back and forth between platforms introduced inefficiencies and confusion in their work. Administrators would actively monitor WhenIwork to manage their fleet of couriers, triage customer service requests that came through ZenDesk, answer phone calls from restauranteurs, and coordinate refunds with payroll administrators via Slack. One administrator who oversaw operations at one of the hubs said they wished they could purchase software from Salesforce to streamline their workflow better, but it was prohibitively expensive. 

Administrators acted as the buffer between stakeholders and Nosh’s technical infrastructure. When couriers disagreed with delivery instructions or assigned tasks, administrators would be alerted to the situation. For example, couriers recounted times when they were assigned a delivery for a restaurant that was far away towards the end of their shift. In this situation, since the couriers wanted to stop working when their shift ended, they contacted a platform administrator via chat and asked them to manually re-assign the order to someone else. Another example of human intervention by platform administrators occurs when restaurants ask them for the same detailed profit and loss reports they received from mainstream platforms. Because the DataDreamers software Nosh used did not support report generation, backend administrators would manually compile these. The limitations of Nosh's technical infrastructure pushed stakeholders to engage in additional invisible labor to make up for shortcomings: 
\begin{quote}\textit{
    ``The point I'm  trying to make is the platform that we currently use is very limited. So there are many things that we can't access, although the  data is there, so we have to kind of find alternatives in order for us  to create visual reports so that we can analyze the data, send it to  our couriers and our restauranteurs, etc.'' (P2)
}\end{quote}

Repair work and invisible labor were essential to making Nosh function properly. At all points in the food delivery process, the technical infrastructure of Nosh was subject to breakdowns. Stakeholders throughout the food delivery process intervened when the technical systems broke down, relying on human intervention to rectify the situation.

\subsection{Human intervention was necessary to overcome breakdowns}

Human operators in Nosh overcame breakdowns through invisible but essential labor. This labor is what the other stakeholders like because they want to interact with people and not machines when things go wrong. However, human intervention requires deep local knowledge and money, and takes time. 

Human mediation plays a critical role in Nosh's operational logic. Platform administrators regularly intervene during the course of food delivery to provide emotional and informational support to restauranteurs, couriers, and customers. As restauranteurs started re-opening and became busier with dine-in traffic, couriers often had to wait long periods to pick up their orders. This wait time was compounded by the possibility that restauranteurs did not use the Nosh interface correctly to time the order preparation. To defuse potentially tense situations between restauranteurs and couriers, Nosh administrators 
\begin{quote}\textit{
    ``actually created these templates to send to our couriers that are acting out, getting frustrated with restauranteurs ... say[ing] hey we know you're frustrated, but we're all a team here let's work together,  here are some ways that you can you know, deal with those types of situations, you  can walk away, you can contact us, you can you know, take a deep  breath.'' (P3).
}\end{quote} 
These interactions aimed to show couriers that Nosh stood behind them and wanted to support their well-being on the job. Couriers appreciated these interactions, often explicitly calling out the unique degree of the personal connection they had with administrators during our interviews.

A healthy skepticism of the capabilities of the technology used further motivated stakeholders to engage in and rely upon human mediation. Couriers developed rich mental models of what food delivery looked like from the restaurant perspective and anticipated where breakdowns might occur, then worked to address those failure points before problems arose. For example, couriers knew that the restaurant-facing UI was complicated and often led to mistakes with timing so they would call up restauranteurs directly and inquire about order status: \begin{quote}\textit{
    ``So I would literally call every single time every restaurant like ‘I have this order. Here's the order number/ here's the name. Anything else you need? Can you let me know if this food has been received by you and do you have any idea how long it's going to take?'. And obviously, you know, whoever's answering, like, the hostess or whoever, they're just going to say it's going to take this long, and then you have to, like, multiply it by three you know, because they don't really know.'' (D3)
}\end{quote}
Many couriers we spoke to learned how to discern whether the person on the phone worked front or back of the house and knew they had to speak with someone from the back of the house if they were to get an accurate estimate for their order.

Administrators also knew that the algorithmic systems underlying the food delivery process were subject to breakdowns. They treated the algorithmic systems they used proactively, anticipating that they would break down and require human intervention. As one administrator put it: \textit{``our job as dispatchers is to skim and identify issues.''}(P1). Administrators played a vital role in manually checking and correcting the logic of the algorithmic system when complications common to food delivery occurred. This contrasts sharply with mainstream platforms, which have little flexibility for situations deviating from the algorithm's prescribes. For example, a courier described a situation where they were delivering multiple orders, but a restaurant had forgotten to start the timer on their end, introducing delays in the process. The courier knew that waiting for the order to be completed would result in the order they had already picked up getting cold, potentially displeasing a customer. The courier contacted Nosh's dispatchers and informed them of the situation. The dispatcher, understanding that these problems arise in food delivery manually changed the delivery routing system to allow the courier to make a single delivery without being penalized for not picking up the delayed order. Platform administrators are trained to intervene manually in the order dispatching system because they know that problems, such as the one described above, do occur.

Stakeholders act as the glue that holds the technical infrastructure of Nosh together. Platform administrators played a key role in maintaining the cohesiveness of the system: they engage in crucial repair work that takes the form of human-to-human interactions to mediate and patch over breakdowns. Restaurateurs, couriers, and customers point to these interactions as integral to their perception of being respected and having agency during food delivery. However, platform administrators articulate the limitations of their technology when it comes to supporting the business operating at a larger scale. As one administrator put it: \textit{``you will break the business because there's no way that you can  support the current infrastructure in a big city it's just not gonna work.''} (P2). This highlights a major tension in designing within the platform economy: at scale, human-to-human interactions that promote sentiments of respect and agency are often untenable.


\begin{table*}[h]
\begin{tabular}{lp{6cm}p{7cm}}
\toprule
\textbf{Stakeholder Group } & \textbf{Primary Technologies Used} & \textbf{Repair Work Performed} \\ \midrule
Restaurants & Nosh Tablet & NA   \\ \midrule
Couriers & DataDreamers App, WhenIWork, Cell phone & Communicating with restaurants and customers \\ \midrule
Customers  & Nosh Mobile App   & NA  \\ \midrule
Platform Administrators & 
DataDreamers app, WhenIWork, ZenDesk, GSuite, Slack, Cell Phone, Computer & 
Dispute Resolution\\ \bottomrule
\end{tabular}
\caption{Technologies Stakeholders Interact With and Repair Work Performed}
\label{tab:repairwork}
\end{table*}

\section{Discussion}
Our findings show that (1) infrastructural breakdowns are inevitable in platform-mediated food delivery, (2) human intervention is essential to address these breakdowns, and (3) stakeholders' perceptions of agency---the ability to interact, reason, and negotiate with a human representative---is intimately linked to the presence of human intervention. When breakdowns occur, human actors must intervene to repair the food delivery process. However, these acts of human intervention are treated normatively differently depending on the platform \textit{type}. Under mainstream platforms such as DoorDash and UberEats, human intervention is punished and viewed as extraneous to the algorithmically prescribed process. Indie platforms like Nosh, in contrast, view human intervention as an inherent part of the food delivery process and rely upon it. Human intervention allows for two-sided communication between the platform and stakeholders, integral to an increased sense of \textit{agency}. Restaurants, couriers, and customers highly value retaining and exercising agency in an algorithmically mediated system. Stakeholders are willing to join a smaller, less well-known, and less technologically sophisticated indie platform that supports human intervention and allows them to retain and exercise agency. 

In this section, we discuss and extend our findings. The first subsection demonstrates why breakdowns are inevitable in the platform economy and how humans intervene to engage in repair work. The second subsection discusses why human intervention remains undervalued and unseen by mainstream platforms. We conclude by demonstrating how business models can be used as an analytical tool to help designers effectively assess whether their recommendations will work in practice to support human intervention in the platform economy. We show that the tension between scale and human-centered design is a navigable one by drawing attention to the role business models play in shaping infrastructures that support human intervention. However, we demonstrate that addressing this tension requires a shift in thinking about economic and organizational facets of the infrastructure underlying the platform economy.

\subsection{Breakdowns are inevitable in food delivery platforms}
As our findings demonstrate, during the course of platform-mediated food delivery, breakdowns are guaranteed to occur: unexpected traffic jams block routes, restaurants overlook details of orders during rush hour, and customers receive cold food. Systems of coordinated workflows fail due to a disparity between how they are designed and how they are actually used in practice, leading to inevitable breakdowns~\cite{grudinWHYCSCWAPPLICATIONS,leszczynski2020glitchy,lord2022critical}. In the platform economy, algorithms are used to orchestrate work processes, such as matching the supply of couriers with consumer demand. However, algorithmic systems require that work processes be simplified, standardized, and codified~\cite{qadri2022seeing}. When infrastructure designers simplify work processes, they make value-laden decisions about which features are extraneous to the design of their system~\cite{pipek2009infrastructuring}. Mismatches between users' and designers' values can result in breakdowns~\cite{rice2006handbook,pipek2009infrastructuring}. For example, an engineer in San Francisco working in routing algorithms may not anticipate that heavy snowfall in Denver occasionally causes road closures.
Regarding unexpected road closures, the couriers we spoke to improvised alternative delivery routes. However, as D2 noted, these deviations were punished by DoorDash. Despite the importance of human intervention in making platform-mediated work functional, designers often fail to consider \textit{human intervention} as a relevant factor in design process~\cite{chandra2017market,gray2019ghost,qadriPlatformWorkersInfrastructures2021,salehi2015we}. By examining breakdowns within the food delivery process, we were able to identify the invisible, but essential human labor that goes into repairing breakdowns.

Infrastructure scholarship shows us that the `re-conceptualization of breakdowns as normal'~\cite{dye2019if,jackson2012repair} can help surface and recenter the labor, values, and power underlying a given system. In Nosh, systems constantly broke down. However, rather than viewing these points of failure as abnormalities, platform administrators and couriers treated them as an inherent and expected feature of food delivery. In particular, platform administrators discussed how they constantly monitored the algorithmic systems for failure, and couriers discussed how they anticipated human error arising on behalf of the restaurants. These two stakeholder groups integrated acts of repair work into their everyday navigation of Nosh's infrastructure. In doing so, they not only maintained but also \textit{transformed} Nosh's infrastructure in a way that preserved humanistic values and gave it the flexibility to function in an unpredictable and idiosyncratic context~\cite{jackson2014RethinkingRepair}. Nosh recognized the \textit{value} of this repair work and relied upon it to make their platform functional. However, our stakeholders expressed that this repair work was unrecognized under mainstream food delivery platforms and punished as a deviation from algorithmically prescribed instructions. Given the importance of repair work in platform-mediated food delivery, why do mainstream platforms fail to recognize and value this process while indie platforms design for it? We argue that this failure can be traced back to mainstream platforms' operational logic, which prioritizes automation and scalability.

\subsection{Human intervention remains undervalued and unseen in the mainstream platform economy}
The presence of human intervention played a significant role in addressing breakdowns during the course of food delivery. Despite human intervention's essential role in making the platform economy viable, it is often unseen and undervalued by mainstream platforms~\cite{elish2020repairing,gray2019ghost}. Drawing on critical labor and feminist critique, we hypothesize that human intervention remains unsupported in mainstream platforms because their `methods of seeing'~\cite{qadri2022seeing} preclude the inclusion of contextual and high-resolution data and their `scale thinking'~\cite{hanna2020against} over-simplifies complex work processes.

In Nosh, platform administrators proactively monitored the dispatching algorithm and communicated with couriers throughout their shifts. Nosh was able to \textit{see} the landscape of food delivery through the couriers' eyes, allowing the administrators to manually adjust algorithmically prescribed instructions when reality didn't align with predictions. By incorporating couriers' vision into the platform's operation, Nosh was able to foster couriers' sense of agency and maintain the flexibility of their infrastructure. In contrast, mainstream platforms rely on ``large observational datasets, which are granular but low-context and low-resolution''~\cite{qadri2022seeing}. These datasets fail to capture the nuances of the landscape that couriers perceive, thus constraining the platform's vision. Additionally, couriers have limited ways of communicating directly with administrators at mainstream platforms. Because mainstream platforms' methods of seeing overlook the rich, local knowledge needed to support gig work, the human intervention that makes platforms functional often remains overlooked.

Mainstream platforms not only overlook the importance of human intervention, but they also undervalue it. Many mainstream platforms fall prey to `scale thinking'---the unwavering commitment to growth through scalability~\cite{hanna2020against,selbst2019fairness}. In the context of algorithmically mediated work, this type of thinking can result in the design of infrastructures that fail to account for local complexity~\cite{qadri2022seeing,hasinoff2022scalability}. Algorithmic code designed for scale makes simplifying assumptions that fail to account for the complexities of gig work, such as unexpected road closures and human error when preparing a complicated order. As we saw with D2, scale thinking can lead to breakdowns when edge cases that are unaccounted for arise. Unlike algorithmic code, platform administrators can respond to and address unforeseen edge cases. Despite its low-fidelity infrastructure, Nosh effectively addressed breakdowns by integrating human intervention and expertise into its infrastructure. Involving users in infrastructure design is ``a pragmatic approach to ... manage issues of complexity and standardized interfaces between new and existing work infrastructure.''~\cite{pipek2009infrastructuring}. Yet, mainstream platforms often fail to account for, integrate, and value the local expertise of stakeholders.

Neglect of local expertise and human intervention in large-scale algorithmic systems can be traced back to computational disciplines prioritizing universal technology design~\cite{jacksonPolicyKnotReintegrating2014,qadri2022seeing}. In the context of gig work, mainstream platforms attempt to supplant local expertise with algorithmic systems: Human labor is seen as something to automate away, a hopefully fungible cost~\cite{iraniDifferenceDependenceDigital2015,doornPlatformCapitalismHidden2020}. However, the view that data-intensive algorithmic models can solve fundamental human problems is misguided: integrating human expertise into algorithmic systems makes them more robust in practice. Stakeholders identified how they relied upon their local expertise to navigate the food delivery process in mainstream and indie platforms. Yet, the value placed on local expertise differed by type of platform. Within Nosh, human intervention was supported and heavily relied upon;  within mainstream platforms, human intervention was seen as an error and punished. By examining what interventions are supported, we can surface underlying value systems and the factors that inform them~\cite{shestakofsky2020making}.

\subsection{Business models enable and constrain the design of infrastructures}
Mainstream platforms do not ascribe value to the labor stakeholders perform in their acts of repair, even though these acts make the platform economy viable. We contend that platforms' business models inform their values which work to enable and constrain infrastructures that support human intervention~\cite{shestakofsky2020making,hasinoff2022scalability}. Mainstream platforms' business models prioritize scalability to secure funding from venture capitalists and subsequent valuation from shareholders~\cite{graham2012startup,sullivan2016blitzscaling}. The business imperative to scale drives the development of algorithms that can ``expand without having to change itself in substantial ways or rethinking its constitutive elements''~\cite{hanna2020against}. To demonstrate how business models inform value systems and subsequent infrastructure design, we discuss how three different design recommendations, while empirically informed by our findings, would fail in practice in the context of mainstream platforms. We then conclude with a call for researchers to examine the economic and organizational features of the platform economy when attempting to design to support human intervention.

\subsubsection{Design recommendations to support human intervention  fail under mainstream business models.}
\textbf{Making Room for Courier Discretion:} We found that Nosh couriers rely on their knowledge of local geography and familiarity with local restaurants to perform their jobs. Couriers highlighted that the ability to exercise agency during the course of food delivery without fear of retribution was a contributing factor to why they liked working for Nosh more than mainstream platforms. Therefore, food delivery platforms should avoid threatening drivers for reasonable deviations from the prescribed route. In mainstream platforms, allowing for driver discretion would introduce complexity into their fleet management systems which oversee thousands, not dozens, of couriers, and would obstruct their ability to prevent gaming behavior by couriers. Essentially, the scale at which mainstream platforms operate constrains the efficacy of this design recommendation. Additionally, each local Nosh hub hires couriers through an interview process, while mainstream platforms offer a ``log on and drive'' model, limiting the ability to verify courier expertise and capability which contributes to the defensive design of their algorithmic management systems~\cite{rahmanInvisibleCageWorkers2021a}.

\textbf{Promote Humans-in-the-Loop in Delivery Operations:} Restaurants, customers, and couriers valued the presence of human platform administrators who intervened when systems broke down. The human agents in Nosh were helpful to the other stakeholders because they were locally situated: each spoke had its own staff of platform administrators that assisted customers, restaurants, and couriers operating within that locale. Thus, these platform administrators could provide relevant and responsive support that understood the contexts in which other stakeholders were operating. In contrast, mainstream food delivery platforms operate nationally and do not have regional hubs staffed with local support agents to support the commerce of each hub~\cite{beharBuildingUnifiedChat2022,GrubhubEasyCustomer}. Providing the localized support that restaurants, customers, and couriers value would necessitate mainstream platforms decentralizing their operations and establishing regional hubs.

\textbf{Build Systems That Support Transparent Two-sided Dispute Resolution:} Restaurants we spoke to appreciated the support they received from Nosh when problems arose, characterizing Nosh as straightforward and transparent. In contrast, they characterized their experiences with mainstream platforms as frustrating and one-sided. For Nosh, it was extremely important to maintain restaurant satisfaction \textit{because Nosh was entirely dependent on restaurants for investment}. Mainstream platforms---as we depict in Figure~\ref{fig:financialflows}---do not have the same degree of financial obligation to restaurants \textit{because they have access to venture capital and shareholder investment}. These differences in financial dependence fundamentally shape how stakeholders are treated in both systems. It would be difficult to enforce the implementation of more nuanced dispute resolution processes without strong economic incentives. Reiterating what R2 said: \textit{``... it's not just Nosh's financial success, they're also looking out for each individual  restaurant's success because \textbf{[Nosh is]} those individual restaurants''} (emphasis added).

\subsubsection{A path forward: Broadening the scope of design research to include policy and economic considerations}
HCI scholars studying the platform economy often share the goal of trying to improve the conditions for participants by mitigating the negative impacts of technical systems through re-imagining their design. However, the technical interventions proposed often fail to consider the role of policy and economic incentives in shaping their feasibility~\cite{jacksonPolicyKnotReintegrating2014}. Attempting to improve conditions in the platform economy solely through technical interventions without considering the contexts in which those interventions would operate limits HCI scholars' ability to understand \textbf{\textit{what}} these interventions will fundamentally change: stopping at design leaves many questions unanswered. It is well established that the platform economy, despite its claims of offering liberatory access to on-demand work, further entrenches systems of oppression and marginalization (\textit{e.g.}~\cite{mcmillancottomWherePlatformCapitalism2020}); but what, realistically, are technical design recommendations doing to fundamentally reorganize the structure of the platform economy to avoid predatory inclusion and precarious conditions?

We call for HCI researchers seeking to improve conditions in the platform economy to examine two paths forward:
\begin{enumerate}
    \item \textbf{Consider the ``Policy Knot'' that shapes the efficacy of design recommendations.} As Jackson and colleagues suggest, it is imperative for HCI scholars to think through how policy plays a key constitutive role in shaping the deployment and effects of technology~\cite{jacksonPolicyKnotReintegrating2014}. In the case of food delivery platforms, we call for HCI scholars to examine how policy, such as regulations of the types of funding mechanisms available to co-ops versus limited liability corporations, constrain the viability of alternative \textit{business models} that are necessary to support technical interventions that improve conditions of participation. HCI scholars do not need to be completely reliant on external domain expertise for guidance on how to study policy constraints on organizations. Rather, they can turn to early HCI and CSCW work, such as that done by Susan Leigh Star, Wanda Orlikowski, and Anselm Strauss for methodological guidance on conducting empirical research that incorporates policy and organizational factors as key elements of design investigations~\cite{orlikowskiInformationTechnologyStructuring1991,orlikowskiStudyingInformationTechnology1991,starLayersSilenceArenas1999,straussSocialOrganizationMedical1997}. 
    \item \textbf{Continue to examine alternative models within the platform economy.} As Glen Weyl suggests, it is essential for scholars to consider and contribute to the plurality of options in the marketplace~\cite{weylWhyAmPluralist2022}. In other words, scholars must continue to investigate how we can further diversify the options available for participation rather than settle into the mindset that we must operate only within existing options. In the context of food delivery platforms, we call for HCI scholars to continue investigating how alternative models for food delivery can be viable \textit{and} simultaneously support participant well-being. Specifically, HCI researchers should investigate how subsidiarity business models, like Nosh, function compared to scalability business models, like UberEats and DoorDash~\cite{hasinoff2022scalability}. HCI scholars could also explore how to support subsidiarity models through technical mechanisms for governance and coordination across multiple localized hubs to provide organizational knowledge and resources to their employees. 
\end{enumerate}

Complicating design recommendations within the broader considerations of social, political, and economic contexts is not new to the fields of HCI and CSCW, especially in the context of systems of coordinated work ~\cite{starLayersSilenceArenas1999,straussSocialOrganizationMedical1997,grudinWHYCSCWAPPLICATIONS}. By presenting our design recommendations and complicating them by considering the constraints of business models and economic incentives, we draw attention to the need for HCI scholars to consider how proposed technical interventions will work to fundamentally reshape some of the root causes of negative impacts experienced by users. We urge HCI scholars to consider the role of policy in informing the impacts of design, however well-intentioned that design may be, and encourage them to continue exploring how a plurality of options for participation in the platform economy can be supported through technical design. In the words of Susan Leigh Star and Anslem Strauss, two eminent figures in the field of infrastructure studies within CSCW and HCI, we call for HCI researchers to \textit{``look in every circumstance at how the application affects relations of power and the nature of work.''}~\cite{starLayersSilenceArenas1999}.

\section{Limitations}
Our results might not generalize to other indie platforms because we studied a single platform. This platform's location and idiosyncrasies are likely to influence our insights.  Also,  our participant sample is biased because those who responded to our call for participation felt strongly enough about expressing their opinion that they chose to sit for an hour-long interview.  In the future, we plan to expand this research to hundreds of other indie platforms in the United States and gather latent data from online sources beyond interviews.

\section{Conclusion}

We discovered the essential role that human intervention plays in making food delivery platform users feel respected. Restaurants, couriers, platform administrators, and customers all took the risk of participating in a relatively untested model of indie delivery because they had a greater degree of agency and financial security, making them feel respected in ways they were not in mainstream food delivery platforms. When breakdowns occurred during the food delivery process, stakeholders could understand, establish control over, and shape the situation's outcome. Drivers, restaurants, and customers valued the ability to interact and negotiate with a human representative in a way they could not with the algorithmic systems of mainstream platforms. Human intervention from Nosh platform administrators played a key role in smoothing over breakdowns as administrators acted as a buffer between stakeholders and the technology. Drivers also stepped up to intervene when algorithmic systems broke down and felt comfortable doing so because they knew a human dispatch agent was available to reason with. We argue for a platform infrastructure design that incentivizes human intervention given the inevitability of its occurrence. We highlight business models' role in enabling and constraining the design of infrastructures that support human intervention. Specifically, we show that business models prioritizing scalability are ill-suited to supporting human intervention whereas business models operating as subsidiarities are better able to do so. Thus we call for HCI scholars to return to traditions from the early days of the field to be attentive to the broader social, political, and economic contexts at hand when making design recommendations. 

\bibliographystyle{ACM-Reference-Format}
\bibliography{bibliography}

\end{document}